\documentclass[%
 reprint,
 amsmath,amssymb,
 aps,
]{revtex4-2}

\usepackage{graphicx}% Include figure files
\usepackage{dcolumn}% Align table columns on decimal point
\usepackage{bm}% bold math
\usepackage{physics} % for \dd and other commands
\usepackage{newunicodechar}
\newunicodechar{α}{$\alpha$}
    \usepackage{caption}
\begin{document}

\preprint{APS/123-QED}

\title{Layer-parity-dependent interfacial coupling in Nb$_3$Cl$_8$/graphene van der Waals heterostructures}% Force line breaks with \\

\author{Hansheng Xu$^1$} 
\author{Yuchen Gao$^1$}
\email{gaoyuchen@pku.edu.cn}
\author{Xinyue Huang$^{1,2}$}
\author{Weihanzhang Guo$^1$}
\author{Zhijie Ma$^{3,4}$}
\author{Ziqi Liu$^1$}
\author{Pinfan Gu$^5$}
\author{Kenji Watanabe$^{6}$}
\author{Takashi Taniguchi$^{7}$}
\author{Youguo Shi$^{3,4}$}
\email{ygshi@iphy.ac.cn}
\author{Yu Ye$^{1,8,9}$}
 \email{ye\_yu@pku.edu.cn}
 
\affiliation{%
$^1$State Key Laboratory for Artificial Microstructure \& Mesoscopic
Physics and Frontiers Science Center for Nano-optoelectronics,
School of Physics, Peking University, Beijing, 100871, China
}%
\affiliation{%
$^2$Academy for Advanced Interdisciplinary Studies, Peking University, Beijing 100871, China
}%
\affiliation{%
$^3$University of Chinese Academy of Sciences, Beijing 100049, China
}%
\affiliation{%
$^4$Songshan Lake Materials Laboratory, Dongguan, Guangdong 523808, China
}%
\affiliation{%
$^5$MIIT Key Laboratory of Semiconductor Microstructure and Quantum Sensing Department of Applied Physics,
Nanjing University of Science and Technology,
Nanjing 210094, China
}%
\affiliation{%
$^6$Research Center for Electronic and Optical Materials National Institute for Materials Science 1-1 Namiki, Tsukuba 305-0044, Japan
}%
\affiliation{%
$^7$Research Center for Materials Nanoarchitectonics National Institute for Materials Science 1-1 Namiki, Tsukuba 304-0044, Japan
}%
\affiliation{%
$^8$Collaborative Innovation Center of Quantum Matter,
Beijing 100871, China
}%
\affiliation{%
$^9$Liaoning Academy of Materials,
Shenyang, 110167, China
}%

\date{\today}% It is always \today, today,

\begin{abstract}
Strongly correlated two-dimensional systems provide compelling platforms for investigating exotic quantum phenomena. Niobium chloride (Nb$_3$Cl$_8$), a single-band Mott insulator, exhibits a remarkable out-of-plane polarization in its topmost layer that oscillates with layer parity, manifesting as an odd-even effect. Using atomic force microscopy (AFM) and Kelvin probe force microscopy (KPFM), this layer-parity-dependent polarization can be effectively characterized through surface morphology and potential mapping, enabling the unambiguous identification of different surface phases. We then fabricated dual-gate Hall devices by coupling different surface phases of Nb$_3$Cl$_8$ with monolayer graphene to investigate how the topmost-layer out-of-plane polarization influences interfacial coupling and the resulting transport behavior. Our results reveal significant phase-dependent variations in charge transfer, carrier densities, and hybridization gaps (25.2 meV for Phase 1 and 30.0 meV for Phase 2). Density functional theory calculations corroborate these experimental findings, showing that distinct out-of-plane polarizations in the topmost layer lead to different orbital overlaps and interfacial coupling strengths. These findings highlight the critical importance of surface polarization and orbital orientation in engineering the properties of strongly correlated van der Waals heterostructures. 
\end{abstract}

\maketitle

%\tableofcontents

\section*{Introduction}
Two-dimensional (2D) strongly correlated systems have emerged as a frontier in condensed matter physics, offering unprecedented opportunities to explore exotic quantum phenomena such as high-temperature superconductivity \cite{hts,hsc}, quantum spin liquids \cite{qsl,qsl2}, non-Fermi liquid behavior \cite{nfl,nfl2} and moiré topology \cite{moire,moire2}. Among these, 2D Mott insulators are particularly compelling due to their highly tunable physical properties governed by the delicate interplay between the electronic bandwidth ($W$) and the on-site Coulomb correlation strength ($U$) \cite{mott}. When $U \gg W$, the system undergoes a transition from a metallic state to a Mott insulating state, characterized by the splitting of energy bands into upper and lower Hubbard bands \cite{flatband}. The ability to integrate these 2D Mott insulators into van der Waals (vdW) heterostructures provides a versatile platform for engineering strongly correlated physics, where the interfacial coupling strength plays a pivotal role in determining the emergent physical properties \cite{strongc}.

In our previous work, we demonstrated that coupling a Mott insulator (Nb$_3$Cl$_8$) with itinerant electrons from monolayer graphene (MLG) can successfully engineer heavy fermion behavior in an incommensurate vdW heterostructure \cite{gyc}. This designer platform circumvents the strict lattice-matching requirements typically associated with conventional strongly correlated systems \cite{ahf}, allowing for the realization of a gate-tunable metal-insulator transition and band-selective electron effective mass enhancement. The magnitude of the hybridization gap and the resulting correlated states in such heterostructures are profoundly sensitive to the interfacial coupling strength, underscoring the critical importance of interface engineering in modulating strongly correlated systems \cite{interface}.

Recently, we further discovered that exfoliated few-layer Nb$_3$Cl$_8$ nanoflakes exhibit a remarkable suppression of structural and magnetic phase transitions, retaining their high-temperature $\alpha$-phase down to cryogenic temperatures (2 K). This stabilization, driven by exfoliation-induced strain, preserves the unique two-layer AB stacking sequence of the $\alpha$-phase \cite{hxy}. Crucially, the single-layer $\alpha$-phase Nb$_3$Cl$_8$ lacks inversion symmetry along the $c$-axis, leading to an out-of-plane polarization in its topmost layer that oscillates with the parity of the layer number. This layer-parity-dependent surface polarization makes few-layer $\alpha$-phase Nb$_3$Cl$_8$ an ideal and highly tunable platform for investigating interface-coupling-related strongly correlated physics, as the distinct surface terminations (down or up polarization) are expected to interact differently with adjacent materials.

In this work, we investigate the odd-even effect regulation in an artificial heavy fermion system by constructing dual-gate Hall devices composed of MLG and different surface phases of $\alpha$-Nb$_3$Cl$_8$. Using atomic force microscopy (AFM) and Kelvin probe force microscopy (KPFM), we unambiguously identify the distinct out-of-plane polarizations of the topmost Nb$_3$Cl$_8$ layers, which depend on the layer parity \cite{phafm}. Through systematic magnetotransport measurements and density functional theory (DFT) calculations, we reveal that the different surface polarizations lead to significant variation in the interfacial coupling strength with MLG. This difference manifests as distinct transport behaviors, including variations in charge transfer, carrier densities, and the magnitude of the hybridization gap (25.2 meV versus 30.0 meV). Our findings not only demonstrate the critical role of topmost-layer out-of-plane polarization in regulating strongly correlated heterostructures but also establish a novel pathway for designing and manipulating quantum phases through layer-parity engineering.

\begin{figure*}[htp]
    \centering
    \includegraphics[width=\textwidth]{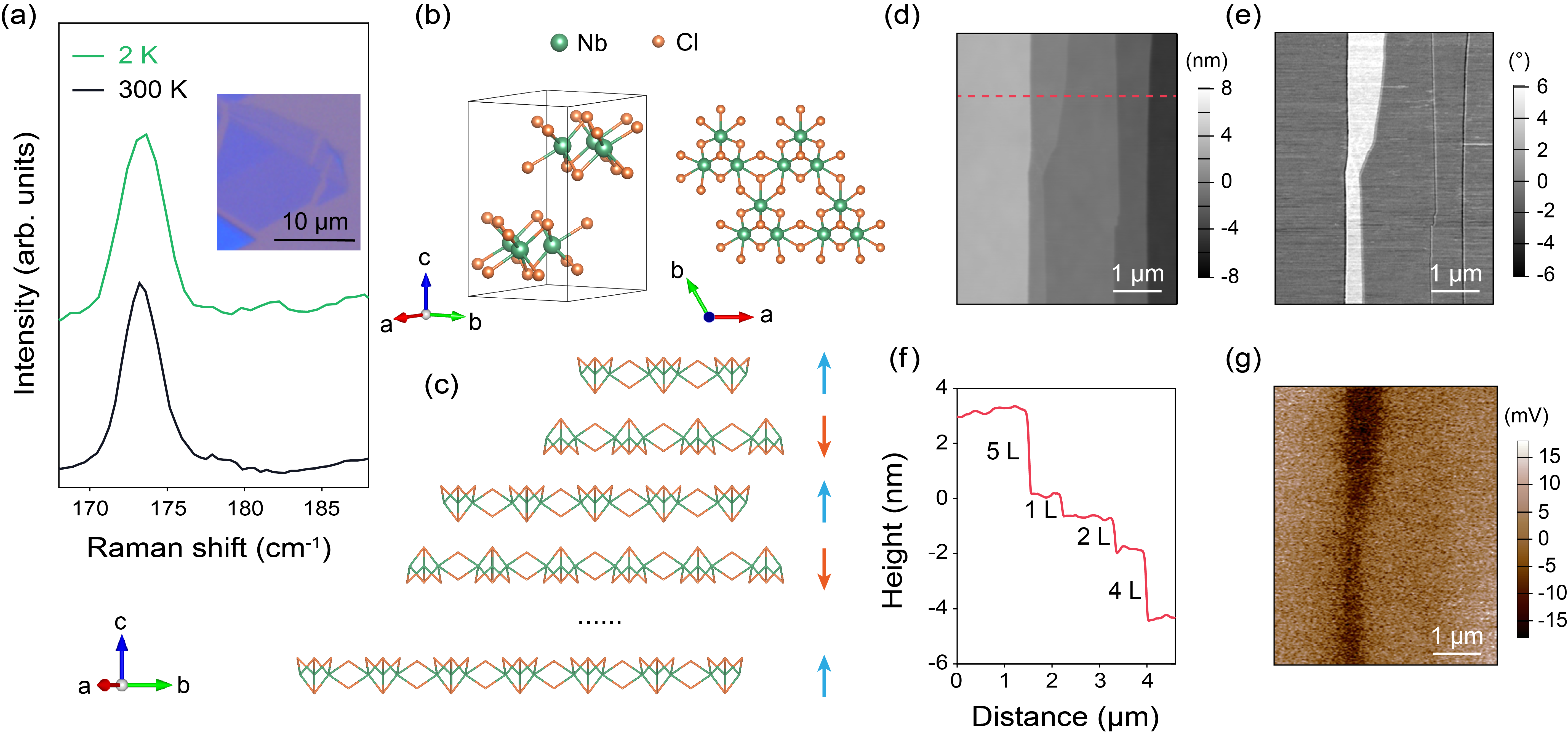}
    \captionsetup{font={small},justification=raggedright,singlelinecheck=false,name={Fig.},labelsep=period}
    \caption{\textbf{Layer-parity-dependent out-of-plane polarization in exfoliated Nb$_3$Cl$_8$.}
    \textbf{(a),} Raman spectra of bulk Nb$_3$Cl$_8$ and few-layer exfoliated samples at 2 K, showing characteristic $\mathrm{A_{1g}^1}$ (at 173 $\mathrm{cm^{-1}}$) peak of $\alpha$-phase. Inset: the optical image of the exfoliated Nb$_3$Cl$_8$ flake. 
    \textbf{(b),} Crystal structure of $\alpha$-Nb$_3$Cl$_8$ showing the kagome lattice with a two-layer periodicity.
    \textbf{(c),} Schematic illustration of the layer-parity-dependent out-of-plane polarization. The different surface terminations lead to distinct electronic properties when coupled with graphene.
    \textbf{(d-e),} AFM morphology (d) and phase (e) maps of a representative few-layer Nb$_3$Cl$_8$ at the same region. 
    \textbf{(f),} Height profile along the marked red dashed line in panel (d).
    \textbf{(g),} The KPFM map of the same region, reflecting layer-parity-dependent surface electric potential.}
\label{F1}
\end{figure*}

\section*{Results}
\subsection*{Layer-parity-dependent out-of-plane polarization $\alpha$-Nb$_3
$Cl$_8$}

Niobium chloride (Nb$_3$Cl$_8$) is a layered two-dimensional Mott insulator. Upon cooling to 100 K, bulk Nb$_3$Cl$_8$ undergoes coupled structural and magnetic phase transitions, transforming from a paramagnetic high-temperature $\alpha$-phase to a nonmagnetic low-temperature $\beta$-phase, accompanied by a rearrangement of its van der Waals stacking sequence \cite{vdW,structure}. The high-temperature $\alpha$-phase crystal structure is shown in Fig. \ref{F1}b. Within each layer, Nb ions form a breathing kagome lattice \cite{bk}. Specifically, three closely spaced Nb ions constitute a Nb$_3$ trimer, and each Nb$_3$ trimer is coordinated by 13 Cl ions, forming a Nb$_3$Cl$_{13}$ cluster composed of three edge-sharing NbCl$_6$ octahedra \cite{mott,38}. Crucially, our previous Raman study revealed that exfoliated few-layer Nb$_3$Cl$_8$ nanoflakes (\textless 100 nm) do not undergo this phase transition upon cooling: instead, they retain their high-temperature $\alpha$-phase down to cryogenic temperatures (Fig. \ref{F1}a), allowing for the investigation of the low-temperature transport of $\alpha$-Nb$_3$Cl$_8$ \cite{hxy}. 

\begin{figure*}[htp]
    \centering
    \includegraphics[width=\textwidth]{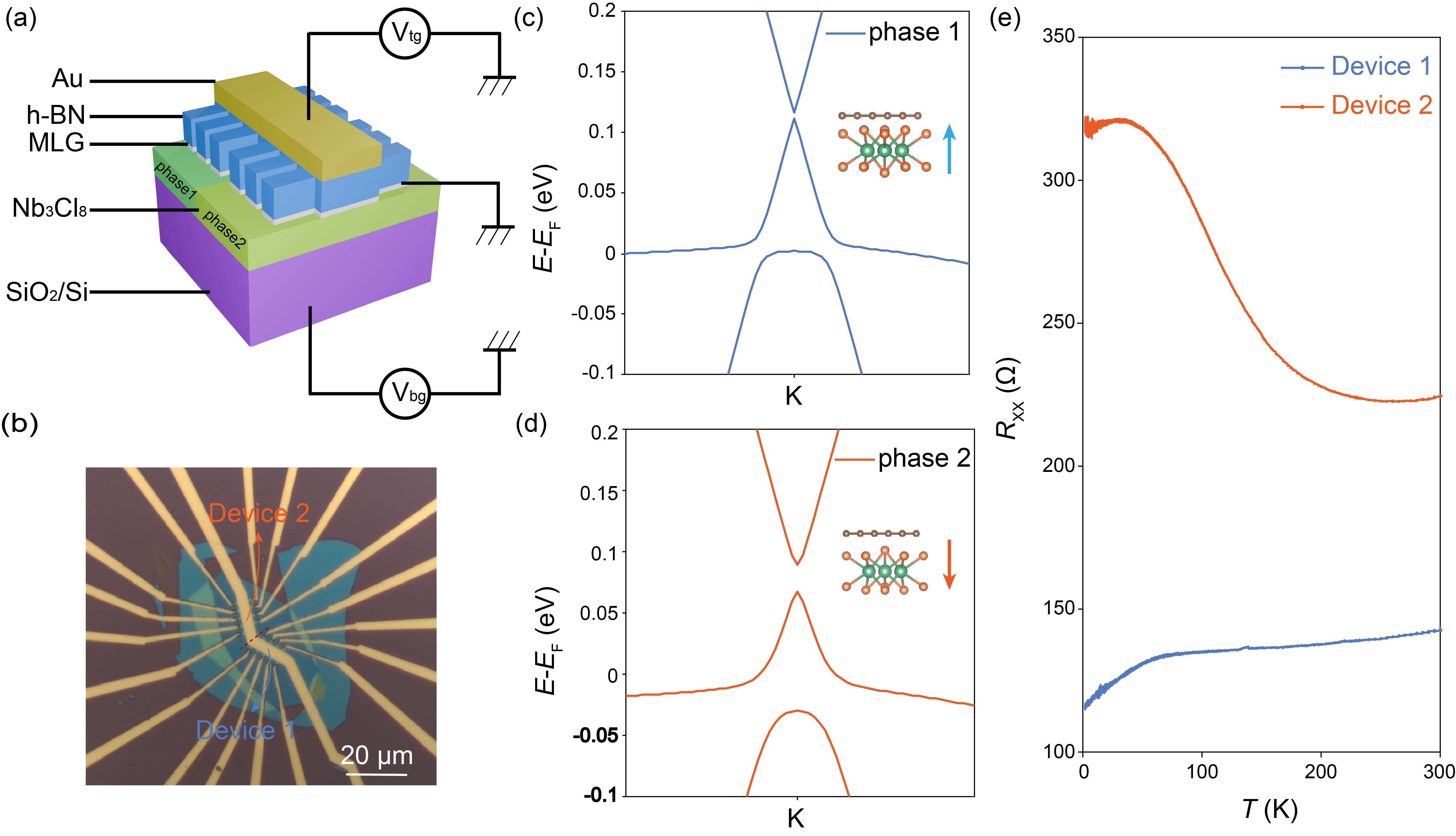}
    \captionsetup{font={small},justification=raggedright,singlelinecheck=false,name={Fig.},labelsep=period}
    \caption{\textbf{Device structure and density functional theory calculations.}
    \textbf{(a),} Schematic illustration of the dual-gated Hall bar device showing the layer stacking geometry. 
     \textbf{(b),} Optical microscope image of the double-gate Hall Device 1 and Device 2 (separated by a red dashed line), in which the MLG is located on a few-layer Nb$_3$Cl$_8$ film with one layer thickness difference. 
     \textbf{(c),} DFT-calculated band structure of the Nb$_3$Cl$_8$/MLG heterostructure for Phase 1 (upward polarization), showing the formation of a hybridization gap.
     \textbf{(d),} DFT-calculated band structure for Phase 2 (downward polarization), displaying a larger hybridization gap compared to Phase 1. Insets show the stacking structures of MLG/different phases Nb$_3$Cl$_8$.
     \textbf{(e),} Temperature-dependent longitudinal resistance of Device 1 and Device 2 at zero gate voltages.}
\label{F2}
\end{figure*}

The $\alpha$-Nb$_3$Cl$_8$ unit cell contains two layers with an AB stacking sequence. Based on the central inversion symmetry of the bulk crystal, the structure exhibits a strict two-layer periodicity \cite{stack}. As illustrated schematically using stick diagrams in Fig. \ref{F1}c, the arrangement of Cl atoms in adjacent layers alternates. Consequently, $\alpha$-Nb$_3$Cl$_8$ lacks inversion symmetry along the $c$-axis within a single layer, leading to an out-of-plane polarization in its topmost layer that oscillates depending on whether the total number of layers is odd or even.

This layer-parity-dependent polarization can be directly visualized using atomic force microscopy (AFM). Fig. \ref{F1}d and Fig. \ref{F1}e display the AFM morphology and corresponding phase images of an exfoliated Nb$_3$Cl$_8$ flake. Fig. \ref{F1}f represents the height profile extracted along the red dashed line in the morphology image. The phase image reveals a striking phenomenon: when crossing a step edge corresponding to an odd number of layers (e.g., 1L or 5L), the AFM phase signal exhibits a stable and significant shift of approximately 6 degrees in the repulsive force mode. Conversely, when crossing a step edge corresponding to an even number of layers (e.g., 2L or 4L), the AFM phase signal remains consistent without any noticeable shift. This distinct phase contrast unambiguously demonstrates that AFM can be used to distinguish between different surface terminations (downward or upward polarization) of the layered sample. This odd-even oscillation phenomenon has been consistently observed across multiple samples (Fig. S1).

Furthermore, Kelvin probe force microscopy (KPFM) measurements reveal distinct surface potential patterns associated with the different out-of-plane polarizations, reflecting the underlying differences in surface charge distribution. As shown in Fig. \ref{F1}g, the surface potential exhibits significant differences between regions with different out-of-plane polarizations. The KPFM results suggest that the distinct surface polarization lead to different local work functions and surface charge distributions \cite{phafm,workfunction}. This difference in surface potential provides a strong prerequisite for expecting different interfacial coupling strengths when these distinct phases of Nb$_3$Cl$_8$ are integrated into heterostructures with other 2D materials, as the differing surface charge distributions and orbital orientations will inevitably affect the interlayer orbital overlap and charge transfer dynamics \cite{ct}. Similar odd-even oscillations have also been detected by scanning tunneling microscopy (STM) measurements \cite{STM}.

\begin{figure*}[htp]
    \centering
    \includegraphics[width=\textwidth]{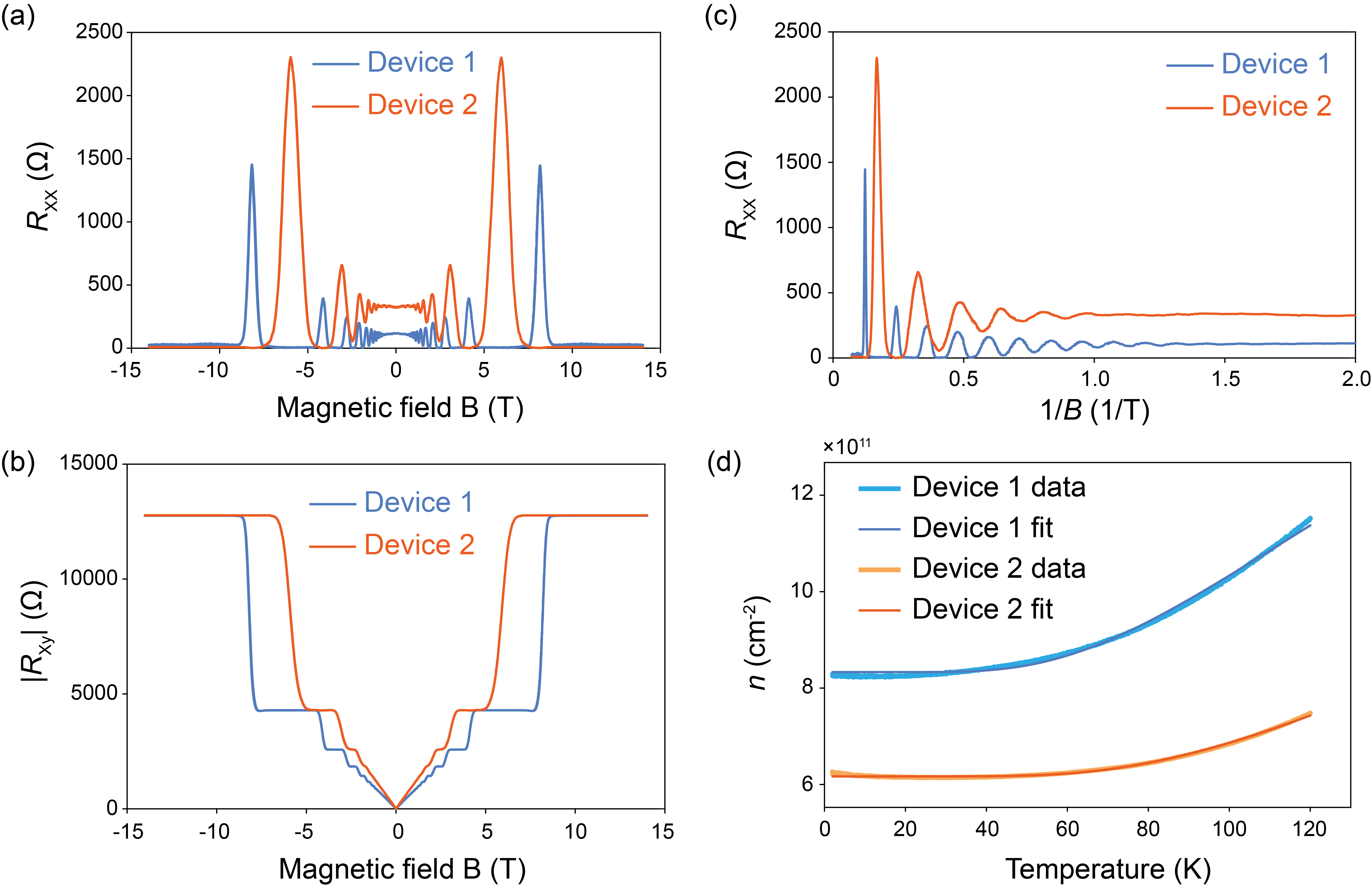}
    \captionsetup{font={small},justification=raggedright,singlelinecheck=false,name={Fig.},labelsep=period}
    \caption{\textbf{Magnetotransport behavior of different phases Nb$_3$Cl$_8$/MLG heterostructure devices.} 
    \textbf{(a),} Longitudinal resistance $R_{\rm{xx}}$ as a function of magnetic field at zero gate voltages, measured at 2 K with the magnetic field applied perpendicular to the sample plane.
    \textbf{(b),} Hall resistance $R_{\rm{xy}}$ versus magnetic field for both devices under the same conditions.
    \textbf{(c),} Shubnikov–de Haas (SdH) oscillations plotted as $R_{\rm{xx}}$ versus $1/B$. The oscillation frequencies yield carrier densities of $n = 8.25 \times 10^{11}$ cm$^{-2}$ ($6.12 \times 10^{11}$ cm$^{-2}$) for Device 1 and (Device 2).
    \textbf{(d),} Temperature-dependent carrier density at $V_{\rm{TG}} = 0$ V extracted from temperature-dependent Hall measurements. The data are fitted using the Fermi-Dirac distribution, yielding hybridization gaps of 25.2 meV for Device 1 and 30.0 meV for Device 2, confirming the phase-dependent interfacial coupling strength. }
\label{F3}
\end{figure*}

\subsection*{Interfacial coupling strength and hybridization gaps}
To systematically investigate how the topmost-layer out-of-plane polarization of Nb$_3$Cl$_8$ regulates the properties of strongly correlated systems, we designed and fabricated vdW heterostructures comprising different surface phases of Nb$_3$Cl$_8$ with MLG. The heterostructures were encapsulated with hexagonal boron nitride (\textit{h}-BN) and fabricated into dual-gate Hall devices. The schematic diagram and a typical optical micrograph of the device are presented in Figs. \ref{F2}a and \ref{F2}b, respectively. We deliberately selected a thin Nb$_3$Cl$_8$ flake featuring a single-layer thickness step (AFM characterization shown in Fig. S1f and Fig. S1g), and fabricated Hall bar devices on both sides of the step for direct comparison. Without intentional twist-angle control, the heterostructure features an inherent lattice mismatch between Nb$_3$Cl$_8$ (6.833 \AA) and MLG (2.468 \AA) \cite{CS}. The two Hall devices corresponding to different phases are designated as Device 1 and Device 2 (Fig. \ref{F2}b). The same metallic top gate and silicon back gate were utilized for both devices to eliminate any variations arising from gate dielectric differences (see Methods).

The KPFM results suggest that the different surface phases of Nb$_3$Cl$_8$ will exhibit distinct coupling strength with MLG. This difference in coupling strength is not merely a consequence of varying interlayer spacing, but is fundamentally rooted in the different orientations of the surface orbitals associated with down and up  electric polarizations, which directly dictate the degree of orbital overlap with the MLG $p_z$ orbitals \cite{pz}. To corroborate this, we performed DFT calculations to evaluate the energy band structures of the different Nb$_3$Cl$_8$/MLG heterostructure phases (Figs. \ref{F2}c and \ref{F2}d). The calculations demonstrate that the Dirac cones of MLG and the flat band of Nb$_3$Cl$_8$ undergo significant energy band reconstruction, resulting in the opening of a hybridization gap. Notably, the hybridization gap opened in the downward-polarized phase (Phase 2) is larger than that in the upward-polarized phase (Phase 1). This indicates a stronger interlayer coupling effect in Phase 2, which can be attributed to a more optimal orbital overlap between the specific surface termination of the up-polarized Nb$_3$Cl$_8$ and the MLG layer (Fig. S2).

The profound impact of the different Nb$_3$Cl$_8$ surface phases is clearly manifested in the transport measurements. Even in the absence of a magnetic field and gate voltage modulation, the temperature-dependent longitudinal resistance ($R$-$T$) curves exhibit striking differences. As shown in Fig. \ref{F2}e, Device 2 displays a pronounced insulating $R$-$T$ behavior at low temperatures, consistent with a well-defined hybridization gap. In contrast, Device 1 exhibits a qualitatively different $R$-$T$ response, with resistance decreasing gradually as temperature is lowered, accompanied by a small downturn at the lower temperatures. These distinct transport behaviors likely arise from the combined effects of two factors: (i) the different interfacial coupling strengths between the distinct surface phases of Nb$_3$Cl$_8$ and MLG, and (ii) the spatial inhomogeneity of the hybridization induced by the lattice incommensurability between Nb$_3$Cl$_8$ and MLG. The smaller average hybridization gap in Device 1 (as predicted by DFT calculations) may result in a more complex interplay between gapped and gapless regions across the device \cite{gyc}. Furthermore, the significantly higher resistance of Device 2 compared to Device 1 indicates a stronger coupling between Nb$_3$Cl$_8$ and MLG, which opens a larger hybridization gap. This will also be discussed in the subsequent part.

\begin{figure*}[htp]
    \centering
    \includegraphics[width=0.8\textwidth]{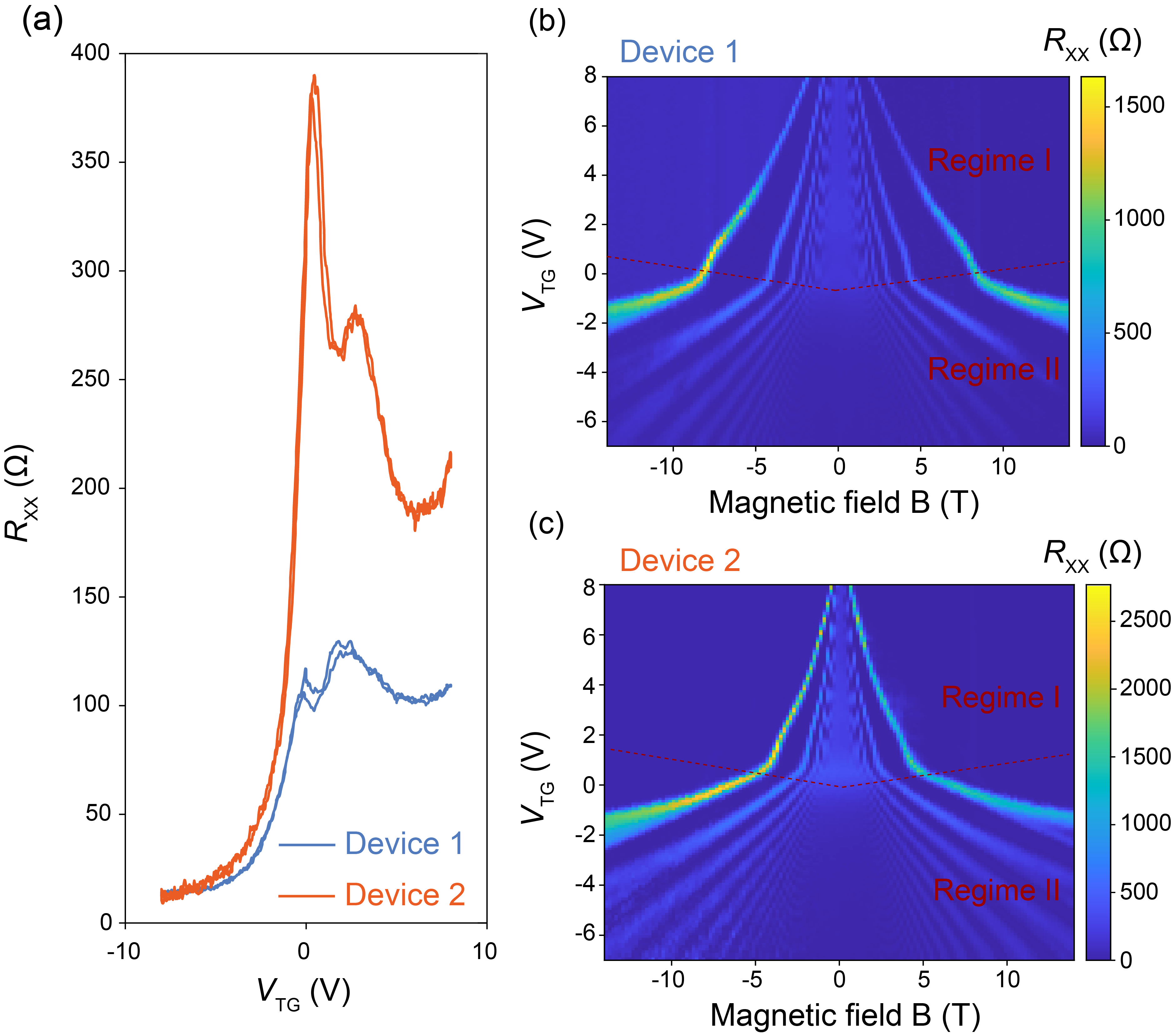}
    \captionsetup{font={small},justification=raggedright,singlelinecheck=false,name={Fig.},labelsep=period}
    \caption{\textbf{Transport behavior of different phases Nb$_3$Cl$_8$/MLG devices under gate and magnetic field regulation.}
    \textbf{(a),} Longitudinal resistance \textit{versus} $V_{\rm{TG}}$ at 2 K without applying a magnetic field, showing distinct gate-dependent behaviors reflecting different interfacial coupling strengths.
    \textbf{(b-c),} Landau fan diagrams showing $R_{\rm{xx}}$ as a function of $V_{\rm{TG}}$ and magnetic field at 2 K for Device 1 and Device 2, respectively. Two distinct regimes separated by red dashed lines are visible. The steeper slope of the Landau fan in Regime I for Device 2 further confirms the larger magnitude of charge transfer, demonstrating the phase-dependent interfacial coupling effect.}
\label{F4}
\end{figure*}

\subsection*{The joint modulation of magnetic field and gate voltage}
To further elucidate how differences in interfacial coupling strength affect strongly correlated transport behavior, we employed magnetic-field and dual-gate-voltage modulation. Magnetic-field scanning measurements at 2 K effectively reveal the transport characteristics of the MLG layer. The longitudinal and Hall resistances as a function of magnetic field (Figs. \ref{F3}a-\ref{F3}b) show excellent quantization, attesting to the high quality of the devices. Notably, Device 1 and Device 2 enter the same quantized plateaus at different magnetic field values, indicating that they possess different carrier densities at zero gate voltage. By analyzing the Shubnikov-de Haas (SdH) oscillations in the longitudinal resistance (Fig. \ref{F3}c) \cite{SdH}, we quantitatively extracted the carrier densities of the two devices. For graphene, where the Landau levels are four-fold degenerate (spin $\times$ valley), the 2D carrier density is given by $n = 4eF/h$, where $F$ is the oscillation frequency. This yields $n = 8.25 \times 10^{11}$ cm$^{-2}$ ($n = 6.12 \times 10^{11}$ cm$^{-2}$) for Device 1 (Device 2) at zero gate voltage. The difference in carrier density directly reflects variations in the charge transfer between MLG and the different surface phases of Nb$_3$Cl$_8$, providing quantitative evidence for the phase-dependent interfacial coupling. The differences in carrier densities will be discussed in detail in the following part.

Temperature-dependent Hall measurements were performed to quantitatively extract the hybridization gaps \cite{Thall}. The data inindicate the presence of thermally activated carriers across the hybridization gap, with the density evolution accurately described by the equation $n = n_0 + n_1 / (1 + \exp(\Delta/kT))$. Here, $n_0$ represents the residual carrier density arising from charge transfer and impurity bands (for Device 1 and for Device 2), $n_1$ denotes the effective density of states at the band edge, and $\Delta$ signifies the excitation gap. As illustrated in Fig. \ref{F3}d, fitting the experimental data yields distinct energy gaps for the different phases: 25.2 meV for Device 1 and 30.0 meV for Device 2. This difference in the thermal activation energy gaps is in excellent agreement with our DFT calculations, confirming that the different surface polarizations lead to different interfacial coupling strengths and, consequently, different hybridization gaps. The extracted effective electron masses (Fig. S3) further reflect the differences in interfacial coupling. Device 2, with stronger charge transfer and interlayer hybridization, exhibits a larger effective electron mass compared to Device 1, indicating a more pronounced band reconstruction effect. These complementary measurements---temperature-dependent resistance, hybridization gap, and effective mass---collectively demonstrate the phase-dependent interfacial coupling in the heterostructure.

We further investigated the relationship between band filling and transport performance by applying top-gate voltage scans under varying magnetic fields. The longitudinal resistance measured at various top-gate voltages (Fig. \ref{F4}a) reveals distinct behaviors for the two devices. Device 1 exhibits a more gradual resistance evolution with gate voltage, while Device 2 shows sharper transitions, reflecting the different magnitudes of charge transfer and hybridization effects. The significantly larger resistance of Device 2 near 0 V also reflects that the hybridization gap is larger due to the stronger interface coupling. The resulting Landau fan diagrams are presented in Figs. \ref{F4}b and \ref{F4}c. Two distinct regions can be identified, separated by a red dashed line. The sudden change in the slope of the Landau energy levels at the boundary between these two regimes indicates charge transfer between MLG and Nb$_3$Cl$_8$ \cite{gyc, transfer, transfer2}. In Regime I, electrons occupy not only the state of MLG but also fill the flat band generated by the hybridization. In Regime II, the Fermi level does not contact the hybridized energy bands, and the electron filling is similar to that of traditional MLG. The different behavior around the phase boundary between the two devices demonstrates that the distinct out-of-plane polarizations lead to different magnitudes of charge transfer and band-structure modifications.
As it crosses the boundary from Region II to Region I, Device 2 exhibits a greater change in the Landau fan slope, causing Device 2 to present a steeper Landau fan slope in Region I. For graphene, the Landau fan slope reflects the ratio of the filling factor to the equivalent gate capacitance: $\mathcal{S} = V_{\rm tg}/ B = \nu e^{2}/(h C_{\rm eff})$, with the effective top-gate capacitance defined as $C_{\rm eff} \equiv e\,\dd n_{\rm MLG}/\dd V_{\rm tg}$. Therefore, a greater slope change indicates a greater change in the equivalent gate capacitance and thereby reflects a larger magnitude of charge transfer through band hybridization and charge filling \cite{gyc, transfer}. This is consistent with the situation corresponding to Device 2, where Nb$_3$Cl$_8$/MLG exhibits a greater coupling.
\subsection*{Microscopic mechanism of phase-dependent interfacial coupling}
The observed differences in transport behavior and hybridization gaps between the two devices highlight the critical role of the topmost-layer out-of-plane polarization in determining the interfacial coupling strength. A more comprehensive understanding requires considering the microscopic orbital interactions at the interface. In the Nb$_3$Cl$_8$/MLG heterostructure, the interfacial coupling is governed by the hybridization between the localized $d$-orbitals of the Nb$_3$ clusters and the itinerant $p_z$-orbitals of the graphene layer. The out-of-plane polarization in $\alpha$-Nb$_3$Cl$_8$ arises from the specific arrangement of the apical Cl atoms relative to the Nb$_3$ trimers \cite{38}. Depending on the layer parity, the topmost surface can be terminated either by an Nb$_3$ cluster with surrounded by more Cl atoms (upward polarization, Phase 1) or more exposed Nb$_3$ cluster (downward polarization, Phase 2). The electron orbital distributions visualized how these different surface terminations affect the spatial extent and orientation of the Nb $d$-orbitals at the interface (Fig. S2). Considering that the role of Cl is to form chemical bonds and fix Nb in the triangular cluster, thereby maintaining the flat band and the conditions for Mott's phenomenon. The 3$p$-orbitals of Cl do not participate in hybridization; instead, the presence of more Cl will shield the hybridization between the Nb $d$-orbitals and the C $p$-orbitals. For Phase 1 (upward polarization), the upper part of Nb is covered by the 3$p_z$ orbital of Cl atoms. The negative charge of Cl repels the 4d electrons of Nb, weakening the vertical extension of the orbitals. This configuration results in weaker interfacial coupling, leading to a smaller hybridization gap (25.2 meV) and less efficient charge transfer. Conversely, for Phase 2 (downward polarization), although the extended Cl atom slightly increases the spatial distance between Nb and C, it does not directly block the overlapping region of Nb 4$d$ and C 2$p$. Therefore, the Nb 4$d$-orbitals are exposed more fully. This enhanced orbital overlap with the graphene $\pi$-band leads to stronger interfacial coupling, which facilitates more efficient charge transfer from the Nb$_3$Cl$_8$ layer to the MLG. Consequently, Device 2 exhibits a larger hybridization gap (30.0 meV) and higher charge transfer between MLG  and Nb$_3$Cl$_8$. 

Transport and DFT results show Device 2 (downward polarization) exhibits stronger interfacial coupling, a larger hybridization gap, and higher charge transfer to MLG at zero gate voltage. Yet Shubnikov-de Haas oscillations reveal Device 1 has a higher measured carrier density---an apparent paradox that reflects the selective nature of quantum oscillations in this multi-hybrid system. In the Nb$_3$Cl$_8$/MLG heterostructure, the lattice incommensurability and wave-function structure of graphene generate spatially nonuniform interlayer hybridization. Consequently, heavy electrons with lower mobility can coexist with residual light electrons with higher mobility. Since the amplitude of SdH oscillations is strongly weighted toward high-mobility carriers, the Lifshitz-Kosevich analysis predominantly probes the residual light-electron channel rather than the heavy-electron channel \cite{LK}. By contrast, the temperature-dependent longitudinal resistance reflects the combined contribution of both channels and is therefore more sensitive to the interaction-induced mass renormalization associated with the hybridized band. The spatial inhomogeneity of hybridization further modulates the phase-dependent coupling. Device 2, with its stronger average hybridization gap, exhibits better spatial uniformity of the gapped region, resulting in a smaller residual carrier density ($n_0$). In contrast, Device 1, with its weaker average hybridization gap, experiences more pronounced spatial inhomogeneity, leading to a larger residual carrier density. The carriers acquire larger effective masses from the flat-band character and experience additional scattering, both of which suppress their SdH oscillation amplitudes. This interplay between the orbital-dependent coupling strength and spatial inhomogeneity naturally explains the observed phase-dependent transport properties.

The back-gate modulation behavior, including the observed hysteresis and bistable resistance states (Fig. S4), will be the subject of a dedicated investigation, as the underlying mechanisms involve complex correlated charge-reconfiguration processes within the Mott insulator that warrant further detailed study.

\section*{Conclusion}
In this work, we systematically investigated the layer-parity-dependent out-of-plane polarization in exfoliated Nb$_3$Cl$_8$ and its influence on interfacial coupling with monolayer graphene. Using AFM and KPFM, we unambiguously identified the distinct surface terminations of different Nb$_3$Cl$_8$ phases and fabricated dual-gate Hall devices to study the resulting heterostructures. Our transport measurements, combined with DFT calculations, reveal that the different surface polarizations lead to significant variations in interfacial coupling strength, manifesting as distinct differences in charge transfer, carrier densities, hybridization gaps, and transport behaviors. Specifically, the downward-polarized phase exhibits stronger orbital overlap with graphene, resulting in enhanced charge transfer and a larger hybridization gap (30.0 meV versus 25.2 meV). These findings establish that surface polarization and orbital orientation are critical parameters for engineering interfacial coupling in van der Waals heterostructures. Our results provide important insights into the design and manipulation of strongly correlated heterostructures through layer-parity engineering.

\section*{Methods}
\textbf{Crystal growth.} 
High-quality Nb$_3$Cl$_8$ single crystals were grown via the PbCl$_2$-flux method. A stoichiometric mixture of high-purity Nb powder (Alfa Aesar, 99.99\%) and NbCl$_5$ (Alfa Aesar, 99.9\%) in a 7:8 molar ratio was ground into a homogeneous blend, loaded into a quartz ampoule, and sealed under vacuum. The charge was first pre-reacted at 700 $^\circ$C for 48 h. The resulting precursor was then combined with sufficient PbCl$_2$ flux, re-encapsulated in an evacuated quartz tube, and subjected to a programmed thermal profile: ramped to 75 ~$^\circ$C over 20 h, maintained at this temperature for 300 h, and subsequently cooled to 500 $^\circ$C over 100 h, followed by natural cooling to ambient temperature. Finally, the excess flux was removed by washing with hot water, affording Nb$_3$Cl$_8$ single crystals of high quality.

\textbf{Characterization of surface polarization.} AFM and KPFM measurements were performed to characterize the surface properties and layer-parity-dependent polarization of the Nb$_3$Cl$_8$ samples. AFM topography and phase imaging revealed the layer-by-layer structure and provided evidence of the odd-even oscillation in surface properties. KPFM measurements recorded the surface potential distribution, which reflects the local work function and surface charge distribution modulated by the out-of-plane polarization. The distinct surface potential patterns observed at step edges with different layer parities provide direct evidence of the layer-parity-dependent polarization.

\textbf{Device fabrication.} Few-layer Nb$_3$Cl$_8$ flakes were mechanically exfoliated onto a silicon/silicon dioxide substrate (285 nm SiO$_2$) using the standard tape-based mechanical exfoliation method in a room-temperature air environment. The thickness and layer number of the exfoliated Nb$_3$Cl$_8$ flakes were determined by AFM in tapping mode using a Cypher S system. MLG and \textit{h}-BN flakes were similarly exfoliated onto separate substrates. The heterostructure was assembled layer-by-layer using the dry transfer method: a poly(bisphenol A carbonate) film on a poly(dimethylsiloxane) (PDMS) imprint template was used to sequentially pick up and stack the flakes. Following assembly, the device was then fabricated into a Hall geometry using standard micro-nano processing techniques, including electron beam lithography, development, reactive ion etching, electron beam evaporation, and lift-off procedures. The resulting Hall device allows for independent control of the top-gate and back-gate voltages.

\textbf{Transport measurements.} Magnetotransport measurements were performed in a Helios$^3$ He insert system equipped with a 14 T superconducting magnet. Standard lock-in measurement techniques were employed to simultaneously measure the longitudinal ($V_{\rm{xx}}$) and transverse ($V_{\rm{xy}}$) voltage components. Four SR830 lock-in amplifiers and two 1 M$\Omega$ resistances were used to provide current excitation. The 10 M$\Omega$ input impedance of SR830 is significantly larger than the measured resistances and thus will not affect the results. For temperature-dependent measurements, all data were collected while the sample was being heated. The heating rate was maintained below 0.5 K/min to ensure that the sample temperature was in thermal equilibrium with the thermometer. The temperature-dependent Hall data were derived through antisymmetrization of the temperature-dependent $R_{\rm{xy}}$ at $\pm$0.5 T.\\

\textbf{Density functional theory calculation.}
The first-principles calculation, based on DFT, was performed using Quantum ESPRESSO v7.0 (pw.x)\cite{QE-2009, QE-2017}. The pseudopotentials C.pbe-n-kjpaw\_psl.1.0.0.UPF, Nb.pbe-spn-kjpaw\_psl.1.0.0.UPF, and Cl.pbe-n-kjpaw\_psl.1.0.0.UPF are from the Quantum ESPRESSO pseudopotential database (http://www.quantum-espresso.org/pseudopotentials). The cutoff energy of the plane wave was set to 120 Ry. The calculation of the commensurate Nb$_3$Cl$_8$/MLG heterostructure utilized a monolayer Nb$_3$Cl$_8$ unit cell and a $\sqrt7 \times \sqrt7$ supercell for monolayer graphene, with the distance between mirrors of the heterostructure exceeding 20 \AA \space to avoid electron hopping between layers. The exchange-correlation potential employed was the generalized gradient approximation (GGA) by Perdew, Burke, and Ernzerhof. The magnetic state in the calculation was set to "non-polarized". The on-site Hubbard $U$ of Nb was not included. The crystal structure of the heterostructure underwent full relaxation, and the residual forces on the atoms did not exceed 1E-4 Ry/Bohr (about 0.003 eV/\AA). The interlayer interaction was described using the "DFT-D2" van der Waals correction. A 30$\times$30$\times$1 (50$\times$50$\times$1) k-space grid was employed in the self-consistent calculation for Phase 1 (Phase 2) configuration. The integrated local density of states (ILDOS) was calculated by integrating over an energy window of -150 to -68.3 meV (-167.2 to -85.5 meV) relative to the Fermi level for Phase 1 (Phase 2) configuration on a 64$\times$64$\times$12 grid.

\begin{acknowledgments}
This work was supported by the National Natural Science Foundation of China (No. 12425402) and the National Key R\&D Program of China (No. 2025YFA1411002 and No. 2022YFA1203902). K.W. and T.T. acknowledge support from the JSPSKAKENHI (No. 21H05233 and No. 23H02052) and World Premier International Research Center Initiative (WPI), MEXT, Japan. 
\end{acknowledgments}
\bigskip
\noindent\textbf{Competing interests}\\
\noindent
The authors declare no competing interests.\\

\bigskip

\bibliographystyle{naturemag}
\bibliography{ref}

\end{document}